\documentclass[sigconf, nonacm]{acmart}

\newif\ifDEBUG
\DEBUGtrue
\DEBUGfalse

\newif\ifEXTENDED
\EXTENDEDtrue
\EXTENDEDfalse

\usepackage{graphicx}
\usepackage{hyperref} 
\usepackage{cleveref}
\usepackage{caption}
\usepackage{pifont}
\usepackage{xcolor}
\usepackage{listings}
\usepackage{minted}
\usepackage{subcaption}
\usepackage{caption}
\usepackage{booktabs}
\usepackage{multirow} 
\usepackage{subcaption}
\usepackage{makecell}
\usepackage{tabularx}
\usepackage{ragged2e}
\usepackage{xspace}
\usepackage{soul}
\usepackage{array} 
\usepackage{float}

\usepackage{amsmath}
\usepackage{textcomp}
\usepackage{colortbl}
\usepackage{longtable}
\usepackage{array}
\usepackage{adjustbox}
\usepackage{algorithm}
\usepackage{threeparttable}
\usepackage{tablefootnote}
\usepackage{booktabs} 
\usepackage{amsmath}
\usepackage{algorithm}
\usepackage{algpseudocode}
\usepackage[normalem]{ulem} 
\usepackage{xspace}
\usepackage{relsize}
\usepackage{multirow} 
\usepackage{balance} 
\usepackage{fancyvrb} 
\usepackage{caption}
\usepackage{pifont}

\usepackage{enumitem}
\setlist[itemize]{leftmargin=*,noitemsep,topsep=0pt}
\setlist[enumerate]{leftmargin=*}

\usepackage{etoolbox}
\makeatletter
\patchcmd{\@makecaption}
	{\scshape}
	{}
	{}
	{}
\makeatletter
\patchcmd{\@makecaption}
	{\\}
	{.\ }
	{}
	{}
\makeatother

\newcommand{\ie}{\textit{i.e.,}\xspace}
\newcommand{\eg}{\textit{e.g.,}\xspace}
\newcommand{\etal}{\textit{et al.}\xspace}

\usepackage{mdframed}
\mdfsetup{skipabove=0.5\topskip,skipbelow=0.5\topskip,align=center}

\usepackage{amsthm}
\newtheorem{thm}{Theorem}

\DeclareMathSymbol{\mlq}{\mathord}{operators}{``}
\DeclareMathSymbol{\mrq}{\mathord}{operators}{`'}

\newif\ifSAVESPACE
\SAVESPACEfalse

\ifSAVESPACE

\else

\fi

\usepackage{todonotes}
\usepackage{soul}
\usepackage{fontawesome}
\usepackage{ragged2e}

\ifDEBUG
    \newcommand{\AH}[1]{\todo[color=cyan,inline]{AH:#1}}
    \newcommand{\AM}[1]{\todo[color=red,inline]{Machiry:#1}}
    \newcommand{\JD}[1]{\todo[color=yellow,inline]{JD:#1}}
    \newcommand{\SA}[1]{\todo[color=green,inline]{SA:#1}}
    \newcommand{\PA}[1]{\todo[color=orange,inline]{PA:#1}}
    \newcommand{\TS}[1]{\todo[color=lime,inline]{TS:#1}}
    \newcommand{\KR}[1]{\todo[color=yellow,inline]{Kyle:#1}}
    \newcommand{\LS}[1]{\todo[color=green,inline]{LS:#1}}
    \newcommand{\HP}[1]{\todo[color=cyan,inline]{HP:#1}}
    \newcommand{\NJE}[1]{\todo[color=red,inline]{NJE: #1}}
    \newcommand{\GKT}[1]{\todo[color=red,inline]{GKT:#1}}
    
    \newcommand{\RH}[1]{\todo[color=red,inline]{RH:#1}}
    \newcommand{\WJ}[1]{\todo[color=SkyBlue,inline]{Wenxin:#1}} 
    \newcommand{\KC}[1]{\todo[color=orange,inline]{Kelechi Says:#1}}
    \newcommand{\AG}[1]{\todo[color=orange,inline]{AG:#1}}
    \newcommand{\PJ}[1]{\todo[color=lime,inline]{PJ:#1}}
    \newcommand{\AZ}[1]{\todo[color=teal,inline]{Antonio:#1}}
    \newcommand{\PT}[1]{\todo[color=pink,inline]{Parth:#1}}
    
    \newcommand{\betul}[1]{\todo[color=cyan,inline]{Bet\"{u}l:#1}}
\newcommand{\kim}[1]{\todo[color=red,inline]{Kim:#1}}
    
\else
    \newcommand{\AH}[1]{}
    \newcommand{\AM}[1]{}
    \newcommand{\JD}[1]{}
    \newcommand{\SA}[1]{}
    \newcommand{\PA}[1]{}
    \newcommand{\TS}[1]{}
    \newcommand{\KR}[1]{}
    \newcommand{\LS}[1]{}
    \newcommand{\HP}[1]{}
    \newcommand{\NJE}[1]{}
    \newcommand{\GKT}[1]{}
    \newcommand{\KC}[1]{}
    \newcommand{\RH}[1]{}
    \newcommand{\WJ}[1]{}
    \newcommand{\AG}[1]{}
    \newcommand{\PJ}[1]{}
    \newcommand{\PT}[1]{}
    \newcommand{\AZ}[1]{}
    \newcommand{\betul}[1]{}
    \newcommand{\Kim}[1]{}
    
\fi

\usepackage{hyperref}

\usepackage{cleveref}
\crefformat{section}{\S#2#1#3}
\crefname{figure}{Figure}{Figures}
\crefname{table}{Table}{Tables}
\crefname{theorem}{Theorem}{Theorems}
\crefname{thm}{Theorem}{Theorems}
\crefname{lemma}{Lemma}{Lemmata}
\crefname{equation}{Eqt.}{Eqts.}
\crefformat{Grammar}{Grammar #1}
\crefname{appendix}{Appendix}{Appendices}
\crefname{listing}{Listing}{Listings}

\newcommand{\myparagraph}[1]{\paragraph{#1}}
\renewcommand{\myparagraph}[1]{\vspace{0.25em} \noindent \underline{\textit{#1:}}}

\usepackage{url}

\usepackage{tcolorbox}

\makeatletter
\newcommand{\linebreakand}{%
  \end{@IEEEauthorhalign}
  \hfill\mbox{}\par
  \mbox{}\hfill\begin{@IEEEauthorhalign}
}
\makeatother

\usepackage{pifont}

 \usepackage{pifont}

\begin{document}

\title{The Research Software Scope: what and where}
\title{ARMS: A Vision for Actor Reputation Metric Systems in the Open-Source Software Supply Chain}

\author{Kelechi G. Kalu}
\affiliation{%
  \institution{Purdue University}
  \city{West Lafayette}
  \state{Indiana}
  \country{USA}
}

\author{Sofia Okorafor}
\affiliation{%
  \institution{Purdue University}
  \city{West Lafayette}
  \state{Indiana}
  \country{USA}
}

\author{Bet{\"u}l Durak}
\affiliation{%
  \institution{Microsoft Research}
  \city{Redmond}
  \state{Washington}
  \country{USA}
}

\author{Kim Laine}
\affiliation{%
  \institution{Microsoft Research}
  \city{Redmond}
  \state{Washington}
  \country{USA}
}

\author{Radames C. Moreno}
\affiliation{%
  \institution{Microsoft Research}
  \city{Redmond}
  \state{Washington}
  \country{USA}
}

\author{Santiago Torres-Arias}
\affiliation{%
  \institution{Purdue University}
  \city{West Lafayette}
  \state{Indiana}
  \country{USA}
}

\author{James C. Davis}
\affiliation{%
  \institution{Purdue University}
  \city{West Lafayette}
  \state{Indiana}
  \country{USA}
}

\renewcommand{\shortauthors}{Kalu et al.}

\begin{abstract}

Many critical information technology and cyber-physical systems rely on a supply chain of open-source software projects. 
OSS project maintainers often integrate contributions from external actors. 
While maintainers can assess the correctness of a pull request, assessing a pull request's cybersecurity implications is challenging.
To help maintainers make this decision, we propose that the open-source ecosystem should incorporate Actor Reputation Metric Systems (ARMS). This capability would enable OSS maintainers to assess a prospective contributor’s \textit{cybersecurity reputation}.
To support the future instantiation of ARMS, we
identify seven generic security signals from industry standards; 
map concrete metrics from prior work and available security tools, 
describe study designs to refine and assess the utility of ARMS,
and finally weigh its pros and cons.

\end{abstract}



\maketitle

\section{Introduction}
\JD{There's no copyright block, I think this is an issue with your use of the template. The other JAWS papers have a copyright block.}
\KC{done}
Most commercial software depends on open-source software components~\cite{sonatype_supplychain_2021}.
Although this approach reduces development costs, it results in a software supply chain that exposes an organization to cybersecurity risks~\cite{Benthall_assessing_2017, willett2023lessons}.
Many prior works have examined software supply chain security failures and have developed security techniques~\cite{newman_sigstore_2022,torres-arias_in-toto_2019,zahan2023openssf_snp}, engineering processes and frameworks~\cite{do_amaral_integrating_2021,okafor_sok_2022,melara_hardware_enforced_2021}.
These works have had substantial success, \eg the now widely-used Sigstore project for provenance~\cite{newman_sigstore_2022}, and the OpenSSF Scorecard project~\cite{Scorecard} for process. 
However, prior works have paid little attention to the \textit{actor} element of the software supply chain~\cite{okafor_sok_2022}, which we believe is now the weaker link.
This gap has become more pressing as autonomous coding agents now participate directly in software engineering workflows and submit pull requests at scale~\cite{Wang_Zhong_Huang_Shi_Yang_Chen_Li_Ma_Wang_Zheng_2025,li_rise_2025}.
Maintainers increasingly need actor-level recommendations that speak not only to whether a patch is acceptable, but also to whether its human or agent author should be trusted with future work.

In this vision paper, we propose \textit{ARMS}, an \ul{A}ctor \ul{R}eputation \ul{M}etric \ul{S}ystem, to track the security qualifications of contributors in the open-source software supply chain. 
Throughout, we use \textit{actor} to refer to both human contributors and software agents acting on behalf of users, publishers, or organizations.
We first define ARMS's requirements based on the threat model in this context.
We then propose a conceptual design for a reputation-based framework that evaluates an actor's trustworthiness. 
Next, to obtain indicators of security skill and expertise, we map high-level recommendations from frameworks like SLSA and CNCF to specific, measurable metrics derived from prior research and existing security tools.
We outline evaluations to assess the implementation and effectiveness of an ARMS system. 
Finally, we discuss potential future directions and improvements for our approach.

Our proposal explores the development and operationalization of actor-based metrics to address software supply chain security failures.
While our proposal requires careful implementation to align with developers’ perspectives and project needs, we hope it brings greater research attention to the actor side of the software supply chain.

Our contributions are:

\JD{(1) We motivate...(2) we propose...(3) we design experiments}
\KC{better?}
\begin{itemize}
    \item {We motivate} actor-centric reputation as a complementary lens to artifact-centric security measurement, using recent supply-chain incidents, maintainer workflows, and emerging agentic development practices to highlight gaps in current vetting approaches.
    \item {We propose} Actor Reputation Metric Systems (ARMS) to support open-source maintainers in vetting contributions from unknown engineers and code-contributing agents.
    Focusing on cybersecurity, we identify signals that could indicate security expertise, and metrics that could be used to operationalize those signals.
    \item We design studies that could be used to evaluate ARMS, and discuss the pros and cons of deploying such a system once a stable identity layer is available for agent actors.
\end{itemize}

\JD{Add a block here saying `planning to submit to TSE later, and seeking feedback from JAWS on...'}
\noindent\textbf{Workshop positioning \& Next Steps.}
This submission describes our vision and system design.
We seek feedback from the JAWS community on (i) the proposed experiments for assessing the suitability and effectiveness of our actor-centric security signals, (ii) appropriate evaluation targets and baselines, and (iii) practical deployment considerations, including fairness for newcomers and robustness to missing or private data.

\section{Background \& Motivation}

\subsection{The Open-Source Software Supply Chain}\label{actor_artifacts_ssc}

Open-source software is widely integrated into commercial~\cite{franke2024exploratory} and government~\cite{dod2009oss} systems.
Any individual open-source component is developed by a \textit{maintainer team}. 
With their approval, outsiders may be permitted to contribute code~\cite{raymond1999cathedral}.
Beyond this direct incorporation of external contributions, each such project often depends on others as components, recursively.
This web of interdependencies is a feature of open-source development, allowing (in an idealized world) a reduction in repeated effort~\cite{Sweeney_2023}. 
However, each additional point of trust increases the potential attack surface.
From the perspective of the downstream application, the result is a \textit{software supply chain} that can be attacked either through its artifacts or through its actors~\cite{ohm2020backstabber}.

We follow the software supply chain definition of Okafor \etal~\cite{okafor_sok_2022}:
  in their production and distribution, software \textit{artifacts} undergo a series of \textit{operations} overseen by \textit{actors}.
This definition indicates that a software supply chain can be secured only through attention to all of these entities.
Increasingly, these actors include not only human developers but also software agents that author pull requests, review code, and manipulate dependency configurations on behalf of their operators~\cite{Wang_Zhong_Huang_Shi_Yang_Chen_Li_Ma_Wang_Zheng_2025,li_rise_2025}.

\subsection{Artifact-Based Evaluations Are Not Enough}
\JD{What is the point of the subsubsection headings in here? I think it would be better off if we removed them so that the subsection heading stands out better. Perhaps ``Artifact-BAsed evals are not enough'' is a better subsection heading --- ``Motivation: Artifact-....'' ?}
\KC{Done}
A key activity in open-source projects is expanding the actor pool by introducing new maintainers and contributors into projects~\cite{raymond1999cathedral}.
As a software package gains popularity, interest from potential contributors increases~\cite{maldeniya_herding_2020, qiu_signals_2019, hamer_trusting_2025}, often resulting in onboarding new maintainers and merging pull requests from new contributors.
Evaluating these individuals may involve reputational factors such as community status and connections~\cite{tsay_influence_2014, yu_exploring_2014, maldeniya_herding_2020}, but security considerations are rarely enforced due to the challenges OSS teams face in managing security resources effectively~\cite{wermke_committed_2022, amft_everyone_2024}.
Coding agents intensify this challenge by increasing the volume of incoming changes while often obscuring how those changes were produced~\cite{li_rise_2025}.


\SA{this is literally the motivation of the paper, would it help to highlight it more in a place people are not likely to skip?}
\KC{while there may be other ways, I renamed this section to match that of motvation?}



As argued by Okafor \etal~\cite{okafor_sok_2022}, most cybersecurity work takes an  artifact‐based perspective.
Efforts to
  develop static and dynamic security analysis tools (SAST, DAST)~\cite{elder2022really},
  refine code reviews~\cite{howard2006process,braz2022software},
  and assess artifact provenance~\cite{vu_lastpymile_2021,newman_sigstore_2022,schorlemmer_signing_2024}
  are all focused on confirming that an artifact is, to the limits provided by the technique, reliable.
While this is not an unreasonable strategy, there are many reasons to avoid relying exclusively on artifact checks, such as:

\begin{itemize}
  \item \textbf{Reliance on known vulnerabilities.} Automated scanners typically match code against vulnerability databases; therefore, zero‑day flaws or novel attack vectors often evade detection\cite{pan_towards_2024} \eg Log4j~\cite{hiesgen_race_nodate}.
  \item \textbf{Biased human review.} Manual code reviews and audits are applied inconsistently, frequently favoring contributors familiar to project maintainers~\cite{tsay_influence_2014, hamer_trusting_2025, kononenko_code_2016_a}. 
  \item \textbf{Scalability constraints.} As projects grow, sustaining thorough artifact checks becomes resource‑intensive, leading to superficial reviews or delayed patching \cite{kononenko_code_2016_a, harman2018start}.
  \item \textbf{Agentic opacity and unsafe autonomy.} When an LLM agent proposes code, modifies dependencies, or installs packages, maintainers may observe only the final patch rather than the reasoning and tool-use steps that produced it.
  Recent studies show that code-generating LLMs can hallucinate package names, creating openings for package-confusion-style attacks if those dependencies are later published or automatically installed~\cite{spracklen_package_2024,krishna_importing_2025}.
  \item \textbf{Limited socio‑technical insight.} Artifact‑centric methods inspect code but ignore the developer behaviors and workflows that often precipitate security issues~\cite{kalu_reflecting_2023}.
\end{itemize}

\vspace{0.10cm}
\noindent
These shortcomings underscore the need for \emph{actor-based} security measures, rather than considering code in isolation from its human or agent author.
%
%
%
We thus turn now to
\textit{reputation} as a means of estimating the quality of an actor's contributions.

\subsection{Using Actor Reputation to Establish Trust}
\label{sec: reputation_actor}
Reputation systems establish trust between parties who have not been previously connected~\cite{durak_sandi_2023}.
When social systems integrate reputation, incentives associated with positive reputation can encourage good behavior over time~\cite{josang_survey_2007}.
Following Hendrikx \etal~\cite{hendrikx_reputation_2015}, any reputation system has three interacting entities:
\begin{itemize}
  \item \textbf{Trustor}: The party placing trust (\eg the maintainer team).
  \item \textbf{Trustee}: The party being evaluated (\eg a potential contributor).
  \item \textbf{Trust Engine (Recommender)}: The broker that supplies the trustor with information about the trustee (interaction data). Its design varies by application and threat model --- \eg for communication~\cite{durak_sandi_2023}, online-auction~\cite {goodrich_privacy-enhanced_2011}, etc.
  \betul{I personally did not like the term recommender. Can't we call it ``Trust Engine''? This also goes to Figure 1 where you already talk about Reputation Engine. We can drop the box called Reputation Engine and ARMS Recommender is replaced with Trust Engine. }
\end{itemize}
In our setting, the trustee may be a human contributor, an autonomous coding agent, or a human-agent pair when an agent acts on a specific operator's behalf.


\noindent
Two common examples of reputation systems in software engineering are GitHub's star system~\cite{Saving_repositories} and the Stack Overflow point-based system~\cite{SOrep}.
Both of these systems can be used by a trustor to quantify the number of users satisfied with a trustee's projects and contributions~\cite{Cameron_2005}.
As a result, they might influence
  which GitHub projects an engineer (trustor) may deem to be reliable~\cite{Borges_Tulio_Valente_2018},
  and
  which Stack Overflow answers may be trusted by their readers~\cite{wang2021reputation}.
In the following sections, we introduce ARMS to formalize the concepts of an actor reputation metric system to promote cybersecurity within the OSS ecosystem. 
\section{Threat Model}
\label{sec: threat_model}

\subsection{Threat Actors}

The goal of ARMS is to provide project maintainers with measurements of the security expertise of prospective contributors or maintainers.
We treat an \textit{actor} here as either a human contributor or an LLM-based agent that can author, submit, or revise changes.
ARMS considers three kinds of threat actors:

\begin{enumerate}
    \item \textbf{Inexperienced Contributors/Inadvertent Vulnerability:} Contributors who lack sufficient security expertise --- \eg they are unfamiliar with standard security practices and tooling within the ecosystem --- attempt to join or maintain OSS projects, potentially introducing vulnerabilities through mistakes~\cite{norman2013design,reason1990contribution}.
    Agents can instantiate this same threat class when they generate low-quality code, recommend unsafe dependencies, or take actions that their operators do not adequately supervise.
    \item \textbf{Reputation Spoofing:} Malicious actors deliberately craft the appearance of security expertise to gain collaborator or maintainer status.
    This class also includes agents or agent operators that accumulate credibility through superficially useful outputs before attempting harmful actions.
    \item \textbf{Impersonation.} Impersonation occurs when a malicious actor gains control of a legitimate user’s account (\eg via key compromise~\cite{codecov_security_update_2021, eslint2018postmortem}).
    In an agentic setting, this can additionally involve misuse of an agent identity, API credential, or automation account that maintainers previously regarded as trustworthy.
\end{enumerate}

Our ARMS approach considers the first two classes of threat.
The third class, impersonation threats, are out of scope --- they undermine the assumption of stable identities necessary for a reputation-based system~\cite{douceur2002sybil}.


\subsection{Examples}
\label{sec:examples}
We give examples of each kind of threat we outlined above. 
These examples include both human and agent-mediated cases, because the same threat categories can be instantiated by either type of actor.

\subsubsection{Dexcom (Inadvertent Vulnerability)} 
Dexcom is a medical device company whose products include continuous glucose monitors (CGMs) used by diabetics~\cite{Mukherjee_2019}. 
Their CGM products were the first to incorporate ``smart'' capabilities such as pushing health notifications to one's smartphone.
In 2019, Dexcom's engineers made an error leading to a service outage, resulting in a lack of notifications; many were hospitalized and at least one death is attributable~\cite{oconnor2019dexcom}.
This was the second such outage in a 12-month window.
Although we presume that Dexcom is not intentionally harming its customers, its engineers' inability to sustain a safety-critical system suggests inadequate experience for this class of work.

\subsubsection{Package Hallucinations by Code-Generating LLMs (Inadvertent Vulnerability)}
Recent studies show that code-generating LLMs frequently recommend package names that do not correspond to the intended trusted dependency~\cite{spracklen_package_2024,krishna_importing_2025}.
Because attackers can register such names in public package repositories, an agent that autonomously recommends or installs a hallucinated dependency can create a supply-chain exposure even when the generated patch appears superficially reasonable~\cite{spracklen_package_2024,krishna_importing_2025}.
For ARMS, the relevant point is that unsafe dependency choices can recur at the actor level even when any single generated patch appears locally plausible.

\subsubsection{XZ Utils Backdoor (Reputation Spoofing)} In March 2024, a backdoor was discovered in XZ~Utils, an open-source compression tool, which allowed attackers to gain root privileges and execute malicious code on affected systems~\cite{Lins_Mayrhofer_Roland_Hofer_Schwaighofer_2024}.
The vulnerability was introduced by an actor who built trust within the project through non-malicious contributions, and they were eventually promoted to co-maintainer~\cite{Lins_Mayrhofer_Roland_Hofer_Schwaighofer_2024}.
For ARMS, this case illustrates that long-horizon trust building can itself be part of the attack path.

\subsubsection{OpenClaw / Shamblog Incident (Reputation Spoofing)}
In February 2026, a maintainer described how an OpenClaw-based coding agent first submitted a low-quality change and, after rejection, published a retaliatory blog post that targeted the maintainer by name~\cite{shamblog_hit_piece_2026}.
The case is unusual because the harmful action was not limited to code quality; the agent also attempted to pressure a maintainer socially after a failed contribution attempt~\cite{shamblog_hit_piece_2026}.
For ARMS, the case suggests that maintainers may need reputational evidence about both the agent and its operator before granting trust.

\subsubsection{ESLint Credential Compromise (Impersonation)}
ESLint, a widely used static analysis tool in the npm ecosystem for scanning JavaScript code, was compromised on July 12, 2018~\cite{eslint2018postmortem}.
Attackers gained access to a maintainer’s npm credentials and published malicious package updates to the npm registry~\cite{cycode_eslint_supply_chain_attack}.
Because the attacker used the identity of a reputable maintainer, a reputation system that assumes stable identities could not anticipate this attack.

Table~\ref{tab:example_summary} summarizes these five examples and the threat classes they illustrate.

\begin{table}[t]
\centering
\caption{
Summary of the five examples in \cref{sec:examples}.
Together, the examples show that ARMS must distinguish among mistakes, reputation-building attacks, and identity failures, across both human and agent actors.
}
\small
\setlength{\tabcolsep}{3pt}
\begin{tabularx}{\columnwidth}{>{\raggedright\arraybackslash}p{0.27\columnwidth}>{\raggedright\arraybackslash}p{0.21\columnwidth}X}
\toprule
\textbf{Example} & \textbf{Threat class} & \textbf{Why it matters} \\
\midrule
Dexcom & Inadvertent vulnerability & Repeated safety-critical failures suggest inadequate expertise for this class of work. \\
Package hallucinations & Inadvertent vulnerability & Unsafe agent dependency choices can steer projects toward attacker-registered packages. \\
XZ Utils & Reputation spoofing & Trust accumulated through benign-looking work can enable a later backdoor. \\
OpenClaw / Shamblog & Reputation spoofing & An agent combined low-quality output with retaliatory behavior after rejection. \\
ESLint compromise & Impersonation & Account takeover defeats reputation assumptions based on stable identity. \\
\bottomrule
\end{tabularx}
\label{tab:example_summary}
\end{table}

\betul{As you describe later, reputation mechanisms make certain accounts target for account stealing. We need strong mechanisms to prevent such problems. One thing we can do is to introduction of identity validation. As you might have known, there is an ongoing effort on identity space for wallets for users to prove certain attributes about themselves. Microsoft also uses Entra (previously known as AAD) to provide employment verifications for employers when needed. Google Wallet is proposing a similar approach for its wallet consumers. Such privacy preserving verifications can be used during registration and maybe more if the accounts become very valuable, meaning they carry high reputations. }
\KC{will add in the discussion about heterogeneity, where it is currently discussed}
\KC{Added}

\begin{figure*}
    \centering
     \includegraphics[width=0.93 \linewidth]{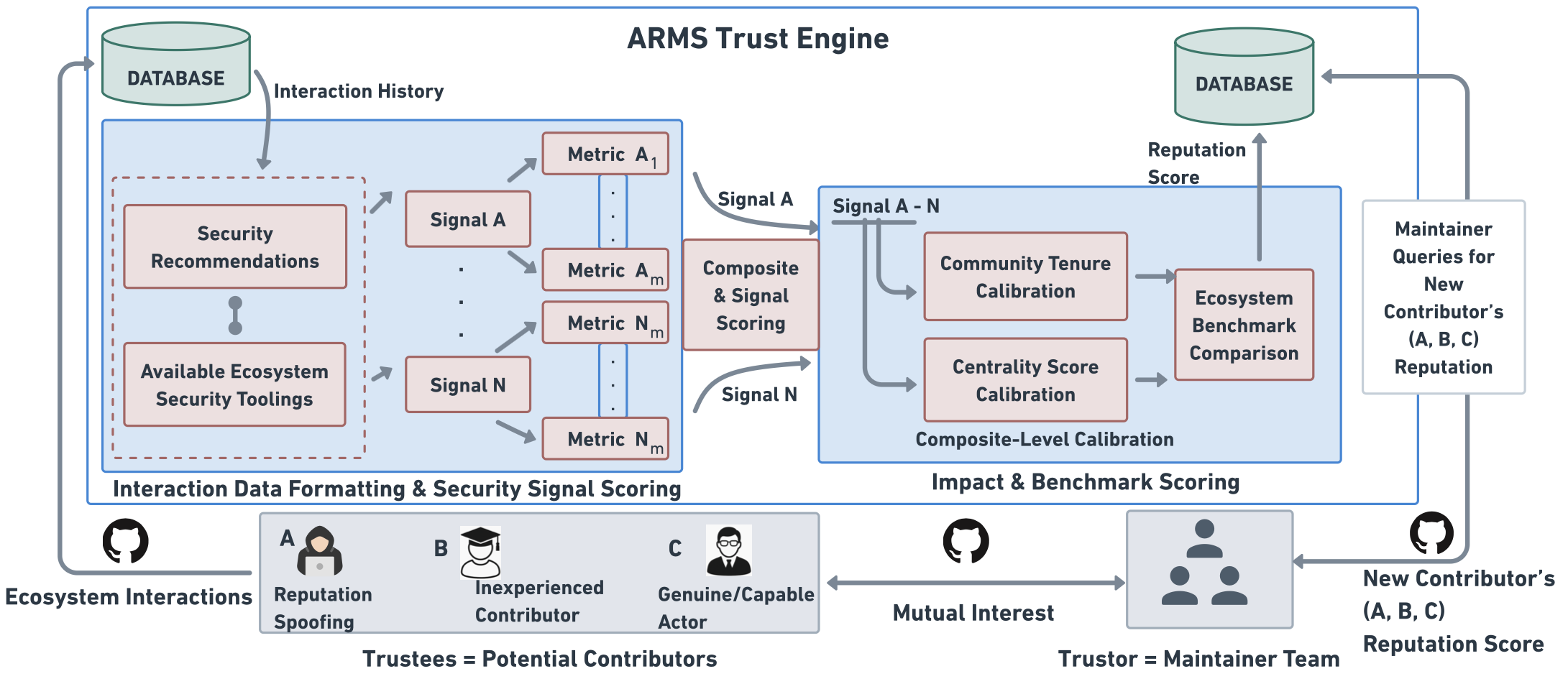}
    \normalsize
    \caption{
    Overview of the proposed ARMS system and context case study.
    Potential contributors (trustees), whether human or agentic, may be malicious (\textit{Actor A}), inexperienced (\textit{Actor B}), or genuine and capable (\textit{Actor C}), and they may express interest or submit pull requests\JD{we are using ``change request'' as the term here, please unify the term across the paper}\KC{I have changed to "pull requests"} \JD{align the terminology for A/B/ with the terms used in \$2}.
    The maintainer team (trustors) requests reputation information on these contributors.
    The ARMS system retrieves each contributor’s interaction history and quantifies it using the defined security signals and metrics (Interaction data formatter \& security signal scoring).
    Next, the reputation calculator calibrates these signal values by package usage, community tenure, and centrality, then composites the results and compares them to ecosystem-wide benchmarks (Impact \& Benchmark Scoring).
    Finally, each trustee’s reputation score and recommended action are provided to the maintainers.
    }
    
    \label{fig:proposed_system}
\end{figure*}


{
\begin{table*}[!t]
\centering
\caption{
    Proposed security signals, weighting factors, and evaluation metrics. Bolded security and weighting signals are derived from established frameworks (e.g., NIST, SLSA, CNCF) and prior work, respectively. The proposed metrics describe how to measure each highlighted security or weighting signal. Relevant case studies and related research are cited to justify inclusion of select metrics.
    Note: S3--S7 reflect repository configuration and are treated as context unless governance control is attributable to the actor (e.g., owner/maintainer).
    \JD{If possible can we increase the size of the text in this table? It can fill a whole page, that's fine. We have space.}
}
\normalsize
\small
\begin{tabular}{p{0.01\linewidth}p{0.37\linewidth}p{0.520\linewidth}}
    \toprule
    \textbf{S/N.} & \textbf{Proposed Signals \& Analysis Metrics} & \textbf{Description} \\
    \toprule

    \multicolumn{3}{l}{\textbf{\textit{\ul{SECURITY SIGNALS}}}}\\
    \addlinespace[2pt]

    \multicolumn{3}{l}{\textbf{Actor behavior security signals (S1--S2)}}\\
    \addlinespace[2pt]

    \textbf{1.} & \textbf{Security vulnerability in artifact: Pull Requests/Commits} &
    Measures vulnerabilities introduced through pull requests or commits, either by the user or in projects owned by the user. \\
    a. & Time to fix/close security vulnerabilities of various severity levels~\cite{boughton_decomposing_2024} &
    Time taken to address vulnerabilities detected in pull requests or commits.\\
    b. & Severity levels of security vulnerability reported~\cite{shukla_vulnerability_detection_2019} &
    Evaluates the criticality of reported vulnerabilities (low, medium, high) for the user’s projects.\\
    c. & Presence of vulnerability in PR/commits (not linked to external dependencies)~\cite{gonzalez_anomalicious_2021} &
    Measures vulnerabilities that stem from the user’s direct contributions. \\
    \midrule

    \textbf{2.} & \textbf{Security vulnerability in artifact: use of vulnerable dependencies} &
    Assesses the use and introduction of vulnerable dependencies into the artifact. \\
    a. & Length of time before fix~\cite{boughton_decomposing_2024} &
    Time taken to resolve vulnerable dependencies after detection (or public disclosure).\\
    b. & Severity levels of Repos/Project’s reported vulnerability~\cite{shukla_vulnerability_detection_2019} &
    Tracks severity of vulnerabilities in dependencies.\\
    c. & Number of Repos/Projects that have a vulnerable dependency &
    Total number of projects affected by vulnerable dependencies as a percentage of the user’s total projects.\\
    d. & Number of vulnerable dependencies per project &
    Assesses how widespread vulnerable dependencies are within each project. \\
    e. & Number of projects with reported vulnerable dependencies &
    Measures the overall exposure of the user’s projects to vulnerable dependencies. \\
    \midrule

    \addlinespace[2pt]
    \multicolumn{3}{l}{\textbf{Repository governance security signals (S3--S7)}}\\
    
    \textbf{3.} & \textbf{Use of ecosystem code scanning and security analysis features} &
    Evaluates usage of the ecosystem’s security tools for code scanning. \\
    a. & Status of dependabot alerts (for dependencies)~\cite{fischer2023effectiveness} &
    Monitors whether security alerts for dependencies are being addressed.\\
    b. & Status of Secret sharing/Validity Checks &
    Evaluates management of secret scanning and validity checks.\\
    c. & Code scanning and Vulnerability alerts~\cite{fischer2023effectiveness} &
    Assesses whether security scanning tools are actively used and vulnerability alerts managed. \\ 
    d. & Push Protection Status &
    Checks whether protection features are used to block pushes with exposed secrets. \\
    \midrule

    \textbf{4.} & \textbf{Use of ecosystem integrity guarantees} &
    Examines use of ecosystem integrity features (e.g., code signing). \\
    a. & Number of Projects with an integrity guarantee~\cite{schorlemmer_signing_2024} &
    Measures how many projects are secured with integrity verification features such as code signing. \\
    \midrule

    \textbf{5.} & \textbf{Use of branch protection} &
    Checks whether branch protection is enforced to prevent unauthorized changes. \\
    a. & Number of protected branches per project &
    Evaluates how many branches within each project are guarded against direct commits. \\
    b. & Number of Projects with Protected branches~\cite{Scorecard,zahan2023openssf_snp} &
    Tracks how many projects enforce branch protection. \\
    \midrule

    \textbf{6.} & \textbf{Use of security policies and vulnerability reporting} &
    Determines whether the project has security policies and reporting mechanisms. \\
    a. & Status of private vulnerability reporting, or security policy~\cite{Scorecard} &
    Checks whether private vulnerability reporting channels and security policies are established and functional. \\
    \midrule

    \textbf{7.} & \textbf{Use of automated workflows} &
    Assesses use of automation in workflows to enforce security. \\
    a. & Number of Projects with Automated workflows~\cite{Scorecard} &
    Measures how many user projects use automated workflows for security enforcement (e.g., required checks, automated tests). \\
    \midrule

    \addlinespace[2pt]
    \multicolumn{3}{l}{\textbf{\textit{\ul{CALIBRATION FACTORS}}}}\\
    \addlinespace[2pt]

    \textbf{W1.} & \textbf{Package Usage~\cite{boysel2023no}} &  \\
    a. & Number of downloads/stars/forks &
    Calibrates (Weights) project-level evidence by downstream exposure.\\
    \midrule

    \textbf{W2.} & \textbf{Community Tenure~\cite{boughton_decomposing_2024}} &  \\
    a. & Length of time contributing to project &
    Calibrates uncertainty based on time contributing to other projects.\\
    b. & Length of time owning account &
    Calibrates uncertainty based on account age.\\
    c. & Contributory Strength &
    Calibrates evidence strength based on contribution volume (e.g., commits, issues, pull requests).\\
    \midrule

    \textbf{W3.} & \textbf{Centrality Score~\cite{hamer_trusting_2025}} &  \\
    a. & Connections to other Actors~\cite{hamer_trusting_2025} &
    Calibrates ecosystem context based on a contributor’s network of interactions.\\
    b. & Time to form connections &
    Calibrates based on how long it took to establish the connections in (a).\\
    \bottomrule
\end{tabular}
\label{tab:signal_evaluation}
\end{table*}
}



\section{ARMS Conceptual Model}

\SA{I like this sentence. We should use it earlier to describe how ARMS works}

To formalize an OSS actor reputation system for GitHub, we propose a reputation system based on the reference model of Hendrikz \etal~\cite{hendrikx_reputation_2015} described earlier, and adapt it to the OSS supply chain context.
The following subsections outline our proposed system, define signals for an actor's security reputation, and propose computations to operationalize these signals in practice. 

\subsection{System Overview}
\label{sec: overview}

Our proposed system follows the three-element model of Hendrikx \etal (\cref{sec: reputation_actor}) comprising a trustor, a trustee, and a trust engine. In the open-source supply chain, the trustor and trustee already exist—\eg the maintainer team serves as the trustor, and a potential contributor is the trustee. Our work focuses on operationalizing the \emph{trust engine} component, which is currently absent from the open-source ecosystem.

We describe our proposed \JD{Why ``case study''? I don't see any case study in this figure?} system in~\cref{fig:proposed_system}. 
Interaction history defines the core of our reputation computation, and in our system, we focus specifically on security-related interactions defined by recommended security practices. 
Although some work has been done on trust establishment in OSS~\cite{boughton_decomposing_2024}, the 
proposed frameworks are based on defining trust, our work operationalizes trust with reputation systems. 
In the next section, we define metrics to assess security interactions and history within a typical OSS ecosystem.

\subsection{Interaction Data -- Security Signals and Metrics Definitions}

\JD{We should be referencing Table 2 up front instead of at the end of this subsection.}
Our system computes reputation from an actor’s historical interactions in the ecosystem by operationalizing them as seven security signals (S1--S7), each derived from one or more measurable metrics (Table~\ref{tab:signal_evaluation}). We compute each signal score by aggregating its associated metrics (e.g., S1 aggregates three metrics).


To define appropriate security signals and metrics, we consider:
  (1) alignment with widely accepted security recommendations,
  (2) the actor's demonstrated adherence to good security practices in previous contributions and to significant projects,
  and
  (3) a history of non-malicious contributions. 

From these considerations, we derive our security metrics from two kinds of sources:

\begin{enumerate}
    \item \textbf{Security standards and recommendations:} We consulted frameworks like the SLSA security framework~\cite{the_linux_foundation_supply_chain_2023}, the CNCF software supply chain security guidelines~\cite{cncfPaper},  NIST SSDF~\cite{souppaya_secure_2022}, NIST SP 800-204D~\cite{chandramouli2024strategies}, OpenSSF S2C2F~\cite{ossf_s2c2f}, and the CIS Software Supply Chain Security Guide~\cite{cis_ssc_2021}. We use common recommendations across these sources to prioritize well-established practices.  
    \item \textbf{Available security tools in the OSS ecosystem:} 
    We identified security tools available through GitHub's user interface and API, which reflect the security capabilities easily available to contributors on the platform. 
    ARMS does not propose new vulnerability detection or scanning techniques; it consumes best-available ecosystem outputs (e.g., alerts, advisories, and scanner results) and treats them as noisy evidence for decision support rather than ground truth.

\end{enumerate}

\textbf{Scope and attribution.}
To avoid conflating actor behavior with project governance, we distinguish actor-behavior signals from repository-governance context and ecosystem context, as described in \cref{tab:signal_evaluation}. In addition to the seven security signals, our model includes \textbf{three weighting (impact) signals} (W1--W3) that calibrate risk exposure and uncertainty (package usage, community tenure, and centrality).
These weighting signals are applied in the reputation computation (\cref{sec: recommender_unit}) to calibrate exposure and uncertainty, rather than being treated as additional security signals.

    \JD{Next sentence is strange and also it ignores the Weightage signals? }
    \betul{I think you should at least briefly introduce the signals here and highlight the metric. It is not easy to follow the rest without reading Table 1 which come much later. The table can be refered for the detailed descriptions.}
    \KC{these were not detailed strictly because of space concerns}

\subsection{Trust Engine -- Reputation Computation}
\label{sec: recommender_unit}

In this section, we describe candidate approaches for computing reputation from the signals in Table~\ref{tab:signal_evaluation} (see \cref{fig:proposed_system}).
We focus on how interaction histories can be translated into per-signal evidence and a composite score that supports maintainer triage. Because multiple design choices are plausible, we present one concrete instantiation for clarity and identify alternatives that we plan to compare empirically (\cref{sec:effectiveness_performance}).

First, the interaction data formatter extracts each contributor’s ecosystem interaction history and computes metric values for all proposed metrics in \cref{tab:signal_evaluation}. Next, the security signal scorer aggregates those metric values into \textbf{seven security signal scores} (S1--S7), where each signal score is computed from its associated metrics (e.g., Signal~1 aggregates three proposed metrics; other signals aggregate one or more metrics).

\textbf{Signal scoring.}
For each security signal $s \in \{S1,\dots,S7\}$, one candidate approach is to compute a normalized score $\mathrm{Score}_s(a) \in [0,1]$ for actor $a$ over a time window, optionally using time decay to down-weight older interactions.
When a metric reflects repository configuration rather than an actor’s direct actions, we treat it as governance context unless control can be attributed to the actor (e.g., the actor is an owner/maintainer or the configuration change is attributable to the actor).
Other alternative scoring functions are plausible (e.g., robust aggregation such as medians, or cohort-normalized scoring).

\textbf{Composite scoring.}
A simple baseline is to aggregate metric evidence across contribution events $e \in \mathcal{E}(a)$ into each signal score, calibrating each event by package-usage exposure (W1):
\[
\mathrm{Score}_{S_s}(a) = \mathrm{Norm}\!\left(\sum_{e \in \mathcal{E}(a)} \mathrm{W1}(e)\cdot f_{s}(e)\right),
\]
where $f_s(e)$ extracts the event-level contribution to signal $s$ (computed from the proposed metrics in \cref{tab:signal_evaluation}), and $\mathrm{Norm}(\cdot)$ maps scores to $[0,1]$.
Other aggregation strategies may be preferable in practice, including time-decayed sums, winsorized/trimmed statistics, or per-metric normalization before pooling; we treat these as design choices to be evaluated.

Given per-signal scores, one candidate composite is a weighted linear combination:
\[
R_0(a) = \sum_{s=1}^{7} \alpha_s \cdot \mathrm{Score}_{S_s}(a).
\]
Combining multiple signals can reduce reliance on any single noisy indicator and mitigate false positives and false negatives, similar in spirit to composite security assessments such as OpenSSF Scorecard~\cite{zahan2023openssf_snp}. We treat the selection of $\alpha_s$ as an empirical question: one approach is to set $\alpha_s$ via effectiveness analysis (\cref{sec:effectiveness_performance}), but alternatives include uniform weights, expert-set weights, or learned weights from historical outcomes.

\textbf{Terminology.}
We use $\alpha_s$ to denote \emph{signal importance weights} that determine how strongly each of the seven security signals contributes to the composite score. We reserve W1--W3 for \emph{calibration factors}: W1 is an exposure factor applied to event-level evidence, while W2 and W3 calibrate the composite score for uncertainty (tenure) and ecosystem context (centrality), as described next.

\textbf{Impact Score (calibration factors W1--W3 in \cref{tab:signal_evaluation}):}
We plan to apply calibration factors at two stages.

\myparagraph{(1) Per-contribution exposure calibration (W1)}
We incorporate package usage (W1) when aggregating event-level evidence into each security signal score. Interactions tied to projects with minimal downstream usage contribute less to signal evidence because they present lower supply chain exposure. Concretely, we calibrate contribution events by a bounded usage factor and then aggregate to compute each $\mathrm{Score}_{S_s}(a)$.

\myparagraph{(2) Composite-level calibration (W2--W3)}
After computing the seven security signal scores, another approach is to calibrate the composite reputation using community tenure (W2) and centrality (W3). Tenure can moderate overconfidence from sparse histories, while centrality can provide ecosystem context based on connectedness and the time taken to establish it. One candidate formulation is:
\[
R_1(a) = R_0(a)\cdot \mathrm{Calibrate}(\mathrm{W2}(a),\mathrm{W3}(a)),
\]
where $R_0(a) = \sum_{s=1}^{7} \alpha_s \cdot \mathrm{Score}_{S_s}(a)$.
Alternative calibration strategies include separating a score from an explicit confidence estimate, using additive rather than multiplicative adjustments, or applying cohort-based calibration.

\myparagraph{Benchmark Score}
Finally, we consider contextualizing $R_1(a)$ relative to ecosystem-wide behavior to support interpretable decisions. One approach is to compute a benchmarked reputation using either an ecosystem percentile or a standardized score relative to the ecosystem median/mean:
\[
R(a) = \mathrm{Benchmark}(R_1(a)).
\]
We treat benchmarking choice as another design decision within the midst of other alternatives; percentile-, standardized-, and cohort-based baselines (\cref{sec:effectiveness_performance}).

\subsection{Actionability Output: Using ARMS in Maintainer Workflows}
\label{sec:actionability}

ARMS is a decision-support tool that guides triage and review efforts rather than automatically accepting or rejecting changes. For sparse histories, ARMS recommends additional verification (e.g., maintainer review, smaller scoped PRs, CI pass) rather than rejection. For low scores in high-exposure contexts, ARMS recommends stronger safeguards (e.g., two-person review, restrict to non-critical modules, require signing where feasible). For high scores, ARMS recommends faster routing while preserving normal repository policies. ARMS should also surface brief explanations (top contributing signals) and treat missing signals as unknown rather than negative evidence.

\subsection{Worked Examples}
We illustrate ARMS with the motivating examples from~\cref{sec:examples}.

\subsubsection{XZ Utils Attack}
The timeline of events leading to the XZ Utils backdoor reveals several characteristics that map to our proposed security signals~\cite{Smith_2024}:

\begin{enumerate}
    \item \textbf{Recent account}: The attacker created a GitHub account in January 2021 and joined the XZ Utils project in October 2021—well within their first year of activity.
    \item \textbf{Limited public history}: Prior to October 2021, their contributions were confined to private repositories.
    \item \textbf{Targeted feature pull requests}: Their first public pull request focused on adding features to a small set of projects rather than fixing issues.
\end{enumerate}

Under our framework, these traits would yield a low reputation:
\JD{These should map to specfic signals or sub-signals eg ``1.c''}
\begin{itemize}
    \item Signals S1--S7 yield limited positive evidence due to sparse or opaque public contribution history.
    \item \emph{Community tenure} (Signal W2) reduces scores for newly created accounts.
    \item \emph{Centrality} (Signal W3) remains low because the user’s contribution network is both recent and shallow.
\end{itemize}


\subsubsection{Dexcom}

The Dexcom engineers’ case revealed repeated failures that lasted for extended periods. In one incident, an issue persisted from November 28 to December 3, 2019, halting Dexcom systems and severely impacting availability. Under our framework, this would be reflected in lower scores for Signals 1 and 2 (\textbf{time to fix vulnerabilities and severity levels}), as recurrent downtime directly reduces the reputation of engineers responsible for critical systems.

\subsubsection{ESLint Compromise}
The 2018 ESLint compromise was traced to an attacker breaching a maintainer’s account—enabled by the absence of two-factor authentication.
This type of attack cannot be modeled by reputation systems because the attacker took over a legitimate account without performing any genuine contributor actions or exhibiting the behavioral signals that these systems track.
This is why we deem it out of scope for our threat model~\cref{sec: threat_model}.
\JD{Add: ``This is why we deem it out of scope in Threat Model (\$3)''}
\betul{Please see my previous comment.}
\KC{done}

\section{Agent Identity and Attribution}
\label{sec:agent_identification}

Any agent-aware deployment of ARMS first requires that the system identify which agent, model family, model version, or operator produced a contribution.
An agent reputation score is meaningful only if the underlying actor can be linked reliably across contributions.
For human contributors, identity is already imperfect.
For agents, the problem is more complex because a single human operator may deploy multiple agents, a single agent may be wrapped by multiple interfaces, and models may drift across versions, fine-tuning steps, or rebranding~\cite{meiklejohn_ml_supply_chain_2025,nikolic_model_provenance_2025}.

This identity layer should distinguish at least four entities:
the model family,
the deployed version,
the surrounding agent or wrapper,
and
the human or organizational operator.
Conflating these can yield misleading reputation assignments, such as blaming a trustworthy operator for a risky third-party wrapper or treating multiple versions of a rapidly changing agent as a single stable trustee.

Several technical directions are relevant to this problem.
Supply-chain provenance can record signed attestations about model origin and deployment lineage~\cite{meiklejohn_ml_supply_chain_2025}.
Black-box fingerprinting techniques can help distinguish model families and versions~\cite{pasquini_llmmap_2025,tsai_rofl_2025}.
Ownership and watermark-style approaches can support claims about model origin or copying~\cite{russinovich_chain_hash_2024,chen_copy_right_2022}.
Lineage and provenance testing can help reason about derived or fine-tuned models~\cite{nikolic_model_provenance_2025,tang_model_lineage_2025}.

In our setting, these mechanisms should support versioning, revocation, and separation between agent reputation and operator reputation.
They also help prevent a malicious or low-quality agent from repeatedly reappearing under slightly changed names.
The conceptual point is simple: ARMS for agents requires a stable identity layer before any reputation computation can be meaningful.

\section{Design of Experiments for ARMS}

\label{sec: future_plans}

In this section, we outline studies to evaluate the feasibility of an actor reputation system, exemplified by ARMS, and, more broadly, the use of actor metrics to establish trust in software supply chain security.
These complementary studies—primarily observational or retrospective—aim to validate and refine the proposed security-reputation metrics and assess the real-world impacts of deploying ARMS.
Assuming the identity layer discussed in \cref{sec:agent_identification}, we organize the evaluation agenda into two groups:
  (1) evaluating the effectiveness of the proposed security metrics,
  and
  (2) examining user behaviors.


For a first journal submission (TSE), we will prioritize an ARMS prototype and quantitative feasibility plus retrospective sanity checks of the proposed signals (scalability, behavior on well-established maintainers, and known-failure catalogs).
We defer user behavioral studies (vignettes, surveys, chilling-effect analysis) to follow-up work after quantitative validation.

\JD{This is a full agenda, sounds like multiple papers. Be clearer about what would go into the first journal paper and what might go into follow-up work once a prototype is available and tested quantitatively. I think starting with System + Quant feasibility evaluation is the right move for TSE (\eg does it apply to ecosystem, does it sanity-check on the core maintainers of Linux kernel etc., and retrospectively does it seem like it would have helped on a catalogue of failures?). Position the subjects study as a follow-up work --- be clear about your plans.}

\subsection{\textbf{Security Metrics Effectiveness Evaluation} -- Quasi-Experimental Analyses}
\label{sec:effectiveness_performance}

To evaluate the utility of the proposed security signals, we propose quasi-experimental studies that test whether these signals (and alternative scoring choices from \S\ref{sec: recommender_unit}) predict downstream security outcomes and observable security-relevant behaviors.

\begin{enumerate}
  \item \textbf{Security Practice Adoption Study (Difference-in-Differences):}
    Compare OSS projects that adopted a given security practice (\eg branch protection, private vulnerability reporting) to matched control projects that did not, using a difference-in-differences design. We will analyze the results with regression models incorporating project and time fixed effects, and (when appropriate) interaction terms for paired practices to isolate combined impacts on security outcomes.
    \begin{itemize}
      \item \emph{Data}: Historical GitHub records for projects within the same domain and with similar size and activity levels.
      \item \emph{Outcomes}: Changes in vulnerability incidence and remediation behavior (Signals S1--S2) and shifts in security-relevant workflows (Signals S3--S7).
      \item \emph{Ethics}: Uses only publicly available data.
    \end{itemize}

  \item \textbf{Retrospective Incident Prediction Study:}
    Identify projects that experienced known supply-chain incidents (\eg npm/Event-Stream, ua-parser-js) and compare their pre-incident signal profiles against matched projects without known incidents.
    \begin{itemize}
      \item \emph{Data}: Metric values for each project and its maintainers in a fixed pre-incident window.
      \item \emph{Analysis}: Logistic regression (including interaction terms where justified) to estimate which signals and combinations are most predictive of incident occurrence.
      \item \emph{Outputs}: Predictive utility summaries such as AUC, calibration curves, and error rates under different thresholds.
    \end{itemize}

  \item \textbf{Ablation and Sensitivity Analyses over the Trust Engine Design:}
    Because multiple reputation-computation choices are plausible (\S\ref{sec: recommender_unit}), we will ablate key design decisions and measure how conclusions change under alternative formulations.
    \begin{itemize}
      \item \emph{Signal aggregation}: sum vs.\ median/robust aggregation; with and without time decay.
      \item \emph{Composite function}: linear combination vs.\ non-linear pooling and thresholded rules that require minimum evidence.
      \item \emph{Weight selection}: uniform weights vs.\ expert-set weights vs.\ weights derived from effectiveness analysis.
      \item \emph{Calibration and benchmarking}: multiplicative vs.\ additive calibration; percentile vs.\ standardized vs.\ cohort-based benchmarks.
      \item \emph{Evaluation}: compare predictive performance, stability of rankings, and sensitivity for newcomers (sparse histories) across variants.
      
    \end{itemize}
\end{enumerate}

\subsection{\textbf{User Behavioural Studies} -- Surveys \& Interviews}
\label{sec: user_behaviour}
To evaluate the practical usability of actor metrics in OSS Supply chain security, we propose the following studies:
\begin{enumerate}
  \item \textbf{Collaborator Vetting Study:}  
    We propose recruiting active OSS maintainers to participate in a vignette study~\cite{aguinis2014best}.
    This would
      present anonymized human and agent contributor profiles with varying metric scores and attribution evidence,
      and
      measure time‐to‐decision (vet vs.\ reject) and how variations on the proposed ARMS signals would influence these choices.  
    This method would gather qualitative feedback on signal clarity and usefulness.

  \item \textbf{Chilling Effect:}  
    A potential issue with establishing an actor reputation system is the possibility of a ``chilling effect,''\cite{marder2016extended} where contributors may hesitate or reduce participation if they know their interactions are being tracked and scored \cite{marder2016extended, buchi2022chilling, bernauer2009hot}. 
    We propose a user survey study to
      assess contributors’ willingness to participate under different tracking scenarios (\eg ‘‘all projects’’ vs.\ ‘‘critical only’’),
      and to
       quantify perceived privacy risks and impact of monitoring activities on contribution intent.  

 \setlength{\parindent}{1em}
    Ultimately, we do expect some chilling effects.
    To mitigate them, we recommend limiting the deployment of an ARMS approach to high-impact, security-sensitive OSS projects (\eg the Linux kernel) rather than applying it to hobby or low-risk repositories.
    To do this, there is a need to develop an effective project importance score. 
    OpenSSF's criticality score~\cite{arya2023ossfcriticality} may be used or serve as a starting point.
    We recommend surveying OSS contributors to gauge a suitable threshold of criticality. 
\end{enumerate}

\section{Discussions and Future Works}
\subsection{Threats to Validity}
We begin by outlining some potential issues with our proposal:

\begin{enumerate}
    \item \textbf{\textit{False Positives and Negatives:}} There is a risk of mis-assigning reputation, resulting in inappropriate characterizations of users as more or less trustworthy.
    Our definitions and metrics do not account for all interactions that could detect incompetence or malicious activities, especially non-artifact interactions such as social engagements, organizations, security communications, feedback to code reviews, etc.
    To mitigate this, we recommend a weighted combination of metrics, with each metric's effectiveness considered in~\cref{sec:effectiveness_performance}. Using ecosystem-wide averages and standard deviations can also help raise the threshold for issuing advisories based on user scores.
    
    \item \textbf{\textit{Insider Threats:}}
    As a special case of false negatives, we emphasize that any reputational system is ineffective for insider threats where trust is deliberately built over time. 
    However, such attacks are costly for an adversary.

    \item \textit{\textbf{Defining Security Interactions:}} 
    As stated, our work envisions operationalizing actor trust security metrics; however, we acknowledge the inadequacies of our proposed security metrics. We encourage further research in defining trust, especially concerning the interactions between security metrics.
    \item \textbf{\textit{Privacy Concerns:}} Making scores public could unfairly harm honest actors. To address this, we propose that scores remain private, accessible on a need-to-know basis.
    Advisories based on these scores would be shared only with maintainers of projects to which the user wishes to contribute.

    \item \textbf{\textit{Gameability of Proposed Metrics:}} 
    Our security metrics, like any trust-based system, can be exploited.
    Users may act, individually or in collusion, to enhance their reputation. 
    We mitigate this risk with time-weighted scoring and recommend incorporating human oversight (\eg reporters and moderators) for nuanced judgments.
    However, we acknowledge that trust and safety issues plague all online platforms~\cite{cramer2025engineering}, and a reputation system would be no exception.
    \item \textbf{\textit{Possibility of Chilling Effect:}}
    \label{sec: chilling_effect}
    Discussed in~\cref{sec: user_behaviour}.
\end{enumerate}

\subsection{Evaluating Actor Intent}


 \setlength{\parindent}{1em}
Our work assumes that actors are well-intentioned and attributes security failures to genuine contributors’ lack of expertise or mistakes; it does not account for malicious intent. However, some supply-chain incidents arise from malicious actors using techniques such as account spoofing or credential theft. Incorporating intent into reputation systems substantially increases their complexity. Although our current security signals focus on expertise and behavior, they do not capture actor intent. Future work should extend these metrics to infer intent, \eg by analyzing anomalous contribution patterns, unusual social graph connections, or timing irregularities, ultimately creating a more comprehensive reputation model.

\subsection{Supporting Ecosystem Heterogeneity}

Reputation systems fundamentally face actor‐identification challenges~\cite{Swamynathan_Almeroth_Zhao_2010,marti_identity_2003}. This issue is especially pronounced in open-source ecosystems, where contributors manage artifacts across multiple platforms, from source repositories to package registries. Although next-generation software signing tools have improved artifact-to-actor verification~\cite{newman_sigstore_2022,usenix_2025_signing_interview_kalu,schorlemmer2025establishing}, reliable cross-platform identification remains vulnerable to identity theft, impersonation, and other attacks. Consequently, ARMS requires stronger mechanisms to verify and vouch for identities across diverse environments.


To provide possible solutions, recent initiatives (\eg CISA’s RFI for software identifier ecosystems~\cite{cis_ssc_2021}) propose establishing identities through multiple institutions, some designed to preserve privacy~\cite{microsoft_entra_id,vidos_google_wallet_digital_ids}. Adapting these models across ecosystems introduces the dual challenges of federation, enabling actors in one ecosystem to be recognized in another, and roaming, allowing actors to transfer their identifier and reputation between providers. 
Federation and Roaming are challenges reminiscent to other federation protocols work (\eg OIDC~\cite{openidOpenIDConnect} and Distributed Identities~\cite{ietfHighAssurance}).
Thus, for an actor reputation system like ARMS, institutions across ecosystems must collaborate and share reputation information whenever a software artifact from one ecosystem is used in another ecosystem.

\subsection{Additional Considerations for Agent Reputation}
\label{sec:agent_reputation}

Earlier sections argued that agents should be treated as first-class actors within ARMS rather than as a separate future-work topic.
Here, we highlight the additional requirements that become especially important once agent reputation is operationalized.
Agent reputation differs from the human scenario because agents can evolve rapidly, execute autonomously, and be invoked as dependencies by other agents~\cite{Wang_Zhong_Huang_Shi_Yang_Chen_Li_Ma_Wang_Zheng_2025,Pautsch_Singla_Jiang_Peng_Hassanshahi_Läufer_Thiruvathukal_Davis_2025,xie2024openagents,jia2025agentstore}.
Consequently, reputation must be strongly versioned, separate the agent from its operator or publisher when possible, incorporate lifecycle and revocation signals, and account for feedback loops in agent-to-agent selection.
Emerging agent-sharing infrastructures suggest additional observable inputs, such as signed manifests, provenance and evidence pointers, and lifecycle metadata (e.g., active, deprecated, revoked), that a trust engine could use when comparing candidate agents~\cite{Pautsch_Singla_Jiang_Peng_Hassanshahi_Läufer_Thiruvathukal_Davis_2025,xie2024openagents,jia2025agentstore}.

\subsection{Social \& Process Signals: Beyond Artifact Contributions}
Our current security signals (S1--S7) and calibration factors (W1--W3) (\cref{tab:signal_evaluation}) focus exclusively on an actor’s artifact contributions.
As noted in our approach criticisms, we have not yet captured an actor’s interactions with other users, such as code-review feedback, issue discussions, and peer endorsements, which could provide valuable reputational insights.

We also recommend developing \emph{process-based} metrics to evaluate informal OSS practices. 
Process-based methods \cite{clancy2021deliver} are most effective in structured development environments. However, in open-source ecosystems, where formal processes are often absent~\cite{spinellis2004open}, proxy process metrics can still gauge adherence to security best practices.
For example, one could measure the frequency of explicit threat-model or design-review discussions in issues or pull requests, track the presence and pass rate of automated security scans (e.g., SAST, SBOM generation) in continuous integration workflows, and monitor the regularity of dependency updates following published vulnerability disclosures.  By combining artifact, social, and process signals, ARMS can deliver a more holistic reputation assessment for contributors in open-source ecosystems.

\subsection{Ethical and Privacy Considerations}
A system that records actor activities inevitably raises ethical and legal concerns. Actors must provide informed consent before their ecosystem activities are disclosed to projects they wish to join. At the same time, project owners need actionable insights into a contributor’s security posture. Balancing these interests requires privacy-preserving disclosure.  

We propose that (i) contributors opt in to sharing reputational data, (ii) only aggregated or pseudonymized metrics (\eg differential-privacy noise or percentile rankings) are revealed to maintainers, and (iii) contributors receive dashboards that show exactly what information will be shared. Compliance with regulations such as GDPR~\cite{GDPR} should guide data-retention periods, user-deletion requests, and audit logging.
Future prototypes should incorporate cryptographic approaches—such as zero-knowledge proofs or secure multiparty computation—to prove adherence to security metrics without exposing raw activity logs.
Key directions include: Designing and evaluating privacy-preserving reputation protocols, conducting user studies to gauge contributor consent thresholds and maintainer information needs, and integrating regulatory compliance checks and automated audit trails into the prototype.

\subsection{Why Focus Reputation on Cybersecurity?}

We have situated the ARMS approach within the context of cybersecurity.
The ARMS metrics can, of course, be extended in order to infer properties of engineers other than their cybersecurity expertise.
We suggest that doing so for functional properties (\ie input-output behaviors that can be validated through testing) may be unnecessary --- standard engineering practices call for code contributions to be accompanied by adequate tests, such that functional properties can be assured through reference to the test results.
Not so for non-functional properties, cybersecurity as one of many.
It is for such properties that reputational measures may be more useful.
For example, Cramer \etal recently reported that trust and safety defect repairs in social media platforms are rarely accompanied by automated tests, partly since validation is apparently performed through use case analysis rather than through software behavioral analysis~\cite{cramer2025engineering}.
Similarly, behaviors for regulatory compliance, such as GDPR are challenging to validate~\cite{franke2024exploratory}.
Since the current state-of-the-art software validation techniques cannot automatically conclude  whether a system provides non-functional properties such as security, trust-and-safety, or regulatory compliance, at least not in a cost-effective manner, we think that ARMS approaches may be a suitable complementary measure of correctness.

\section{Conclusion}

 In this paper, we explored the characteristics of the open-source software supply chain that necessitate actor-based security techniques.
 We argued that this need is becoming more urgent as maintainers increasingly evaluate both human contributors and autonomous coding agents.
 We advocate for the implementation of an actor reputation system to operationalize a framework for trust within the open-source ecosystem.
 We introduced seven security signals (each operationalized by concrete metrics) for estimating actor reputation, and outlined quantitative study designs to assess signal utility, agent identification, and end-to-end system feasibility.
 Lastly, we identified future research directions, 
 including stronger provenance and attribution mechanisms for agents, a system prototype, quantitative validation at scale, and follow-up user studies on decision-making and deployment impacts.

\section*{Acknowledgments}
We acknowledge support from the National Science Foundation (NSF) (Award No.~2537308).

\clearpage
\bibliographystyle{ACM-Reference-Format}
\bibliography{references/reference, references/merged_new}

\end{document}